\documentclass[12pt]{article}
\usepackage{pdproc,epsfig} 

\begin{document}

\renewcommand{\thefootnote}{\alph{footnote}}
  
\title{{\Large\bf Fifty years of Yang--Mills Theories:}
\vskip 0.3cm
{\Large\bf a phenomenological point of view}}
\vspace{0.5cm}
\author{ALVARO DE R\'UJULA}

\address{ CERN,
Geneva, Switzerland\\
 Physics Dept. Boston University\\
 Boston, Mass., USA\\
 {\rm E-mail: alvaro.derujula@cern.ch}}

\abstract{On the occasion of the celebration of the first half-century 
of Yang--Mills theories, I am contributing a personal recollection of how
the subject,  in its early times, confronted physical reality, that is,
 its ``phenomenology". There is nothing 
original in this work, except, perhaps, my own points of view. But I hope
that the older practitioners of the field will find here grounds for nostalgia,
or good reasons to disagree with me. Younger addicts may learn that
history does not resemble at all what is reflected in current textbooks:
it was orders of magnitude more fascinating.}
   
\normalsize\baselineskip=15pt

\vspace{1cm}
\noindent
{\it This account is dedicated to the memory of my thesis advisor:
Angel Morales.}
\section{Introduction}

 In the book of which this note is a chapter we are celebrating
{\it Fifty Years of Yang--Mills Theories}. Yang--Mills (YM) theories 
are non-Abelian field theories with an {\it ``isotopic gauge invariance''} \cite{YM}. By now, even some schoolchildren are told that in
a non-Abelian gauge theory the gauge quanta are
``charged" sources, e.g. the gluons of quantum chromodynamics
(QCD), which carry a ``colour" charge \cite{GMFL}.
In this sense, Einstein's general relativity is the first non-Abelian 
gauge theory \cite{Uti}, since gravitons gravitate. We might
 as well be celebrating close to 90 years of Einstein's theory of gravity.
But there is at least one reason to concentrate on 
YM theories: they are {\it ``true theories"}, in 
that they are understood at the quantum level, while gravity is not.

Gerard 't Hooft, who is organizing and editing the commemoratory book
on YM theories, has asked me to cover my views of their phenomenological
aspects. You can see that I have accepted. Why?... that is unclear. To judge
from the list of contributors,
 it is quite possible that I shall be the only one showing any 
real data. The responsibility to deal single-handedly with the nitty-gritty details of 
reality is considerable. I shall make it easier for myself
 by covering only a fraction of
the subject: the early developments, which I witnessed from a very short distance.

%\newpage

The most noteworthy theoretical developments after the work of 
Chen Ning Yang and Robert Mills 
will be covered by others, and some of them
have already obtained the proverbial first class
tickets to the freezing winter of Stockholm. I am referring to the work
of Sheldon Glashow, Steven Weinberg and Abdus Salam on the ``standard"
electroweak YM model \cite{SLG,Steve,AS}, and to the
work of Gerard 't Hooft and Martinus Veltman on why YM theories ---even
with massive gauge quanta--- are {\it true} theories \cite{'t,'tV}. 
My involvement with these early developments was small but not
unimportant: I photocopied Shelly's thesis in the Harvard library
and discreetly sent it (upon request) to the relevant address.
The birth of QCD was more of a collective enterprise, and the subject
of whether it will, or will not, deserve someone's trip
 to Sweden is one of the items of
endless conversation in which idle particle physicists indulge.

Let me distinguish the YM electroweak gauge {\it theory} from the
corresponding {\it model}. By {\it theory} I mean the full-fledged
renormalizable construct, as tested ---including quantum corrections---
by the $e^+e^-$ colliders at CERN and SLAC. The no-mean feat of the
experiments was precisely this set of tests, although it is not so easy to ``sell"
their importance to the general public (the Nobel Committee had to work harder
than usual in 1999). By the electroweak {\it model} I mean that part of it
that can be tested at tree level, notably the existence
of weak neutral currents. 

Historically, the {\it phenomenological} aspects of 
the electroweak {\it theory} were less interesting than those of the 
corresponding {\it model} \footnote{Young theorists
often take a cynical view of renormalizability and the distinction I made
between model and theory, being satisfied with ``effective"
theories valid in a very limited energy range. This may be justified in
the case of chiral Lagrangians, but it is harder to justify for YM
theories. And the crucial role that renormalizability played in
the development of QED, QCD and the electroweak theory is
hard to sweep under the rug.}, in the sense that they required less of an
experimental and theoretical
effort to understand what the heck was going on. Also, the understanding
of deep-inelastic electron and neutrino scattering as well as $e^+e^-$
annihilation in the ``charmed" region were intimately intertwined
with the proof of the reality of quarks and with QCD, whose asymptotic
freedom makes it an inescapably subtle quantum theory for starters. 
In practice what all this meant is that particle physics, or at least
its phenomenology, were infinitely more
challenging and interesting in the 70's than they were thereafter.
These arguments are the excuse for the choice of topics I shall discuss
at greatest length (Sections 2 and 3).

\section{Neutral currents}

The 1973 discovery of weak neutral currents in neutrino scattering
by the Garga-melle bubble-chamber collaboration at CERN \cite{Musset} 
made experimentalists, and the world at large, aware of YM theories, 
much as the 1971 work of 't Hooft \cite{'t} immediately attracted attention
from (field) theorists to the same subject. Both lines of
work \cite{'t,'tV,Musset} were monumental
 in their difficulty, and run against deep prejudices.
In the case of neutral currents, the prejudices had various sources,
among them\footnote{I have refreshed my memory on these issues by 
re-reading the talk by Bernard Aubert in \cite{Aubert}, an interestingly 
uneven book; and their rendering by Peter Galison in \cite{Galison}, a 
thoroughly documented report, that reads like a good novel.
I quote them freely in this section.}:
\begin{itemize}
\item{} 
The very strong limits on their strangeness-changing variant, which
in the 70's were at the branching ratio level of $10^{-6}$ ($10^{-9}$) for 
$K^\pm$ ($K^0_L$) decays \cite{NCL}. 
\item{}
The perception
that the measurement of neutrino-induced weak neutral currents ---at least
in the semileptonic channels with larger cross-sections than the 
purely leptonic ones--- was nearly impossible in its practical difficulty \cite{DonP}.
\item{}
The existence of severe (and incorrect) upper limits on strangeness-conserving
neutral-current processes, such as the one by Ben Lee, stating that
{\it [The results of W. Lee \cite{WLee}] rule out the existence of the neutral current
predicted by Weinberg's model...} \cite{BLee}, or the one by John Bell,
J. L{\o}vseth and Tini Veltman: {\it Thus the ratio of neutral-current
``elastic events'' is less than 3\%} \cite{BLV}.
\item{} 
The fact that neutrino experiments at the time were primarily
designed to look for
sequential heavy leptons and for the Lee--Yang process \cite{LeeYang}
---$\nu_\mu+{\cal N}\to W+\mu+{\cal N}$--- for light $W$'s, but not for neutral currents.
\end{itemize}

Naturally, the neutral-current processes favoured by theorists were the
ones whose cross sections could be calculated with confidence in the
standard model: $\nu_l$, $\bar\nu_l$ elastic scattering on electrons,
whose standard cross-sections were
worked out by 't Hooft as early as  1971 \cite{Gerardnu} (he may also
have been the first to emphasize a trivial but important fact:
a measurement of the weak-mixing angle, $\theta_W$, would
imply a prediction for the ---then--- enormous masses of the $W$ and $Z$).
 By January
1973, the Aachen group of the Gargamelle collaboration had found
a ``picture book'' event, with a single recoiling electron \cite{Glepton}. 
But it was
just one event and ---while it immediately had the effect of putting
Gargamellers even harder at work on neutral currents--- various 
cautions and dangerous backgrounds kept the team from publishing this result
until July, right before they published their work on semileptonic 
neutral currents \cite{Musset}.

In parallel to these developments, theorists were at work on what
experimentalists {\it wanted}: predictions for semileptonic
neutral-current cross sections, and in particular lower limits on them
(the minimization is relative to $\theta_W$). Weinberg worked out limits 
on the elastic channels and on $\Delta$ production \cite{SW1}, while being 
amongst the first to recognize the need for charm and the GIM 
mechanism \cite{GIM} to make the electroweak model compatible
with the suppression of strangeness-changing neutral currents.
Finally, Bram Pais and Sam Treiman \cite{PT} and Mani Paschos
and Lincoln Wolfenstein \cite{PW} produced what experimentalists
{\it really} wanted: results for inclusive channels, which were at the
level of 20\% for the minimal ratio of neutral to charged channels.

Neutral-current events at the level of 1/6 of the total! ---nothing would
seem to a theorist easier to detect. Yet it was far from easy to do it
convincingly; the intricate story is well told by Galison \cite{Galison}
and Riordan \cite{Riordan},  so I shall not try to summarize it here.
The decisive talk by Paul Musset at CERN on 19 July 1973, in which 
he reported the Gargamelle results \cite{Musset}:
\begin{eqnarray}
R_\nu&\equiv& \left({NC\over CC}\right)_\nu=0.217 \pm 0.026\nonumber\\
R_{\bar\nu}&\equiv& \left({NC\over CC}\right)_{\bar\nu}=0.43 \pm 0.12,
\nonumber
\label{Paul}
\end{eqnarray}
was received with enthusiasm by some, complete scepticism by others.
One of the sceptics is a colleague of mine at CERN, Nobel Laureate for the
codiscovery of the second neutrino ---to give you an extra hint. He always
disbelieves new results, which is his way of being right some 99\% of the
time. This time was in his 1\% category, and I believe that my good friend
Paul Musset would have also become a Nobel Laureate ---had he not
died at an untimely age\footnote{Whether others may have shared the prize
I cannot tell, nor can I suggest any names at the risk of making $n-1$ enemies,
with $n$ a large number, for Gargamelle was the first modern ``large"
and hard-to-manage collaboration \cite{Galison}.}.

\begin{figure}[]
\hskip 2truecm
%\vspace*{2cm}
\hspace*{-1.5cm}
\epsfig{file=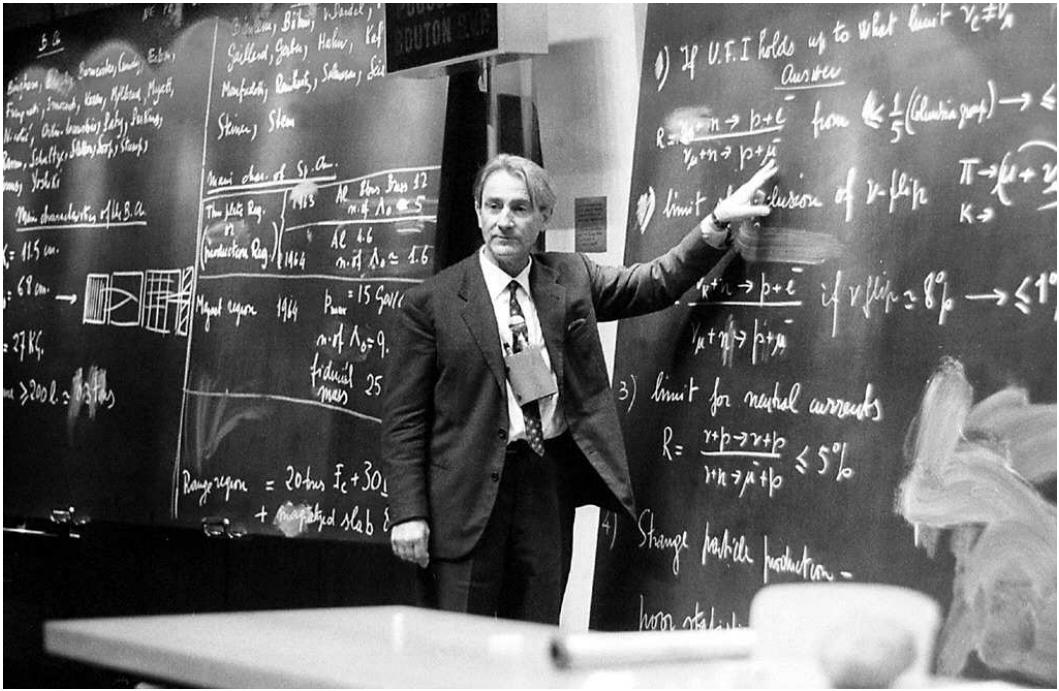,width=14cm}
\caption{Gilberto Bernardini delivering a lecture on neutrino-scattering
results, in 1964. At waist-hight on his left is the wrong result $R\leq 5$\%. Nobody
recalls what he erased further to his left.}
\label{Bernardini}
\end{figure}

Inordinate amounts of spice and emotion were added to the neutral-current 
saga by the ``alternating-current'' results of the HPWF collaboration,
which, after a few cycles,
 finally agreed with Gargamelle \cite{HPWF}. The happy ending
was a talk by Bari Barish at the 1974 HEP conference in London,
in which he reported the neutral-current results of the Caltech/FNAL
collaboration, fully confirming their existence \cite{Bari}.
The difficulty of the early measurements of weak neutral currents can
be summarized in a Nov.~1964 picture of Gilberto Bernardini, reproduced
in Fig.~(\ref{Bernardini}). Notice the upper limit on the ratio $NC/CC$ of ``elastic" 
neutrino cross sections on nucleons... it is significantly below 
the currently established result.

I may have transmitted the impression, which I shall retransmit below,
 that experimentalists may sit
for a long time on a gold mine before convincing themselves that
that is the case. Sometimes theorists are even better at that:

As early as 1961, Gell-Mann and Glashow realized that YM theories were 
endangered by the fearsome spectre of strangeness-changing neutral 
currents \cite{GMG}. Yet, in 1964, 
Hara \cite{Hara} and Bj{\o}rken and Glashow \cite{BjSLG}, 
when suggesting charm and the concomitant
 weak current with the explicit mechanism
for averting the danger, did not realize they had solved the problem
first stated by Gell-Mann and Glashow. Only six years later,
Glashow and/or Iliopoulos and/or Maiani \cite{GIM} must have been idle enough
to read the literature, and realize that the problem and its solution
were stated in papers that at least one of them may have been expected
to have read.

\section{QCD and charm in their early years}

This section concerns the development of QCD
from 1973 to 1978, spanning  the period from the discovery of asymptotic
freedom, covered  by others in this book,  to the discovery of
the gluon\footnote{I make in this section
abundant use of a previous rendering of the same story  \cite{Erice}.}.

\subsection{A Deep Inelastic Dawning}

In a talk reproduced in \cite{HGeorgi} Howard Georgi
recalled how everybody, in years long past, 
knew Harvard as the place {\it not} to be. He was the seventh of a long list
of applicants. The first six had chosen ``better'' destinations. One year later,
I was to share Howard's honour. It turns out that I was not quite at the
right place at the right time, but I was next door.  At the time of the
discovery of QCD's asymptotic freedom  \cite{DDF}, David Politzer had the
office next to mine at Harvard, where he was a Junior Fellow and I a lowly
post-doc. Some of the  outcasts that gathered in this back-door way changed physics (and Harvard) forever. Now Harvard is again {\it a place
where to be}, but for more formal reasons.

In the late '60s, it seemed perfectly ridiculous for the strongly interacting partonic
constituents of protons to do what they do: exhibit a ``scaling''
free-field behaviour  \cite{BjF} in deep inelastic scattering experiments
 \cite{Dis}. Thus, though the rationale for a rather low-energy
``asymptotia'' remained obscure for quite a while, the discovery of
asymptotic freedom was received with a great sigh of relief by a
then-persecuted minority of field-theory addicts.

Knives had been sharpened for long on inclusive reactions. Not
surprisingly, the first concrete predictions of QCD  \cite{moments}
concerned the deviations from an exact scaling behaviour. But the electron
scattering and $e^+e^-$ annihilation data of the time  \cite{Dis} covered
momentum transfers, $Q^2$, of not more  than a few GeV$^2$. Nobody (yet) 
dared
to analyse these data in the ``asymptotic'' spirit of QCD. And that is how
some people not affected by dataphobia ---a morbid condition of the brain
(or brane?) that turns theoretical physicists into mathematicians--- set  out to
exploit the only data then available at higher $Q^2$.

\begin{figure}[]
\hskip 4truecm
%\vspace*{2cm}
%\hspace*{-1.5cm}
\epsfig{file=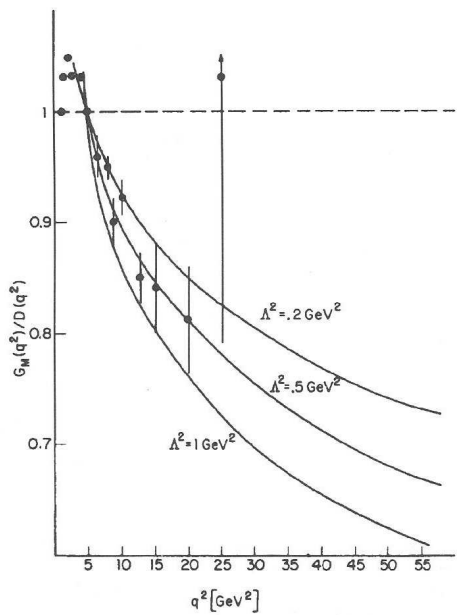,width=7cm}
\caption{The magnetic form factor of the proton normalized to a
dipole fit, a first attempt to measure $\Lambda$ that 
stumbled upon the right result. Thus distilled, the presentation
obscures the fact that the theory and the actual cross-section data
agree in a range of six orders of magnitude.}
\label{Lambda}
\end{figure}

By the early '70s, the proton's elastic form factor had been measured
 \cite{proton} up to $Q^2\sim 20$ GeV$^2$. To bridge the gap between the QCD
predictions for deep inelastic scattering and the elastic form factor, two
groups  \cite{yo,GT} used (or, with the benefit of hindsight, slightly abused) the
then-mysterious ``Bloom--Gilman duality''  \cite{BG} relating the deep
``scaling'' data to the elastic and quasi-elastic peaks. I prefer the paper
containing Fig.~(\ref{Lambda}) and beginning:  {\it ``Two virtues of asymptotically 
free gauge theories of the strong interactions are that they are {\bf not}
free-field theories and they make predictions that are {\bf not}
asymptotic''}; to conclude {\it ``The results agree with experiment but are
{\bf not} a conclusive test of asymptotic freedom.''} 

Lustra later, looking back at the papers I just quoted
 \cite{yo,GT}, I notice that they were received by the publisher on
consecutive days. This was not atypical of the Harvard/Princeton
 competition of those times, a competition
that I then viewed as fierce and {\bf evil}. Mollified by Chronos, I
now view this tug of war as fierce and useful. Being a bit more
inclined to data analysis than my Princeton competitors,
I attempted to measure $\Lambda$, as in Fig.~(\ref{Lambda}), while they
simply chose the ``right'' value... somewhat surprisingly since
their results ---based on an analysis slightly different from mine---
neither fitted the data nor subtracted from their confidence in the theory
\cite{GT}.

\begin{figure}[]
\hskip 2truecm
%\vspace*{2cm}
%\hspace*{-1.5cm}
\epsfig{file=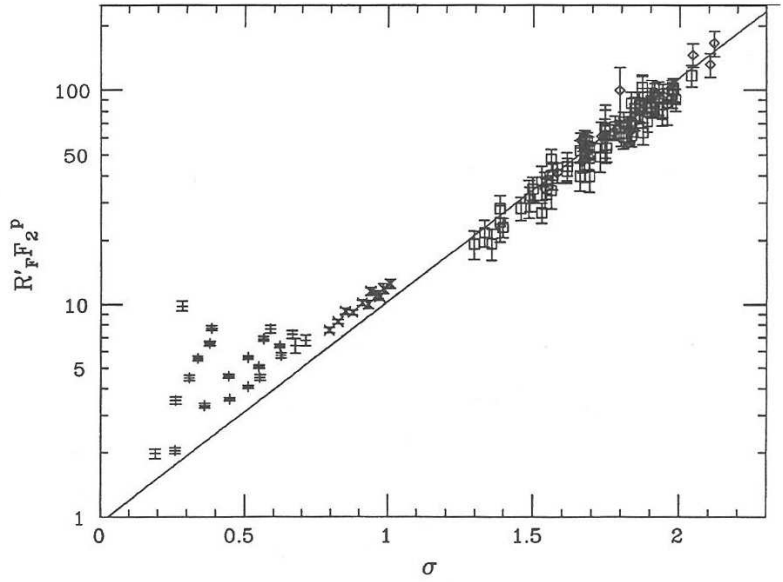,width=11cm}
\caption{The data snuggle toward the slope predicted by QCD,
for $n_f=4$, $N_c=3$.}
\label{Lowx}
\end{figure}

There were sporadic truces in the Harvard--Princeton wars.
An example is a 1974 paper signed by an even mix of six
authors  from the two institutions  \cite{six}. We derived the leading
QCD predictions for structure functions $F(x,Q^2)$ in the ``Regge''
domain, $x\rightarrow 0$. In the consuetudinary notation
and with $R$ an explicit function of $x$ and $Q^2$:
\begin{eqnarray}
&&{\partial \, \ln \, (R \, F)\over \partial \, \sigma} \;
\mathrel{\mathop\rightarrow_{{\rm ``large"}\;\sigma}  } \;
{12 \over \sqrt{33-6\, n_f / N_c}} \; ,\nonumber\\
&&
\sigma^2\equiv s\;\ln {x\over x_0}\;\;\;\; s\equiv \ln { \ln\,
(Q^2/\Lambda^2)\over \ln\, (Q^2_0/\Lambda^2)}\; .
\label{sigma}
\end{eqnarray}
Unbeknownst to many,
this paper enjoyed a second life as the Mother of All Physics at
HERA. The comparison  \cite{BF} of its predictions with experiment,
shown in Fig.~(\ref{Lowx}), 
was a belated but striking confirmation of QCD.

It was not easy to write the Harvard--Princeton paper  \cite{six}. In those
fax-less and email-less
days we had to exchange drafts by post. Many references had two
entries, a ``Harvard'' one and a ``Princeton'' one. The Orangemen would  send a
draft with all references in the P/H order, whence the Crimson would make
some cosmetic changes in the manuscript and send it back with references
reversed. The paper was published the  Princeton way,  which proves
that we  at Harvard were only joking.

\begin{figure}[]
\hskip 2truecm
%\vspace*{2cm}
%\hspace*{-1.5cm}
\epsfig{file=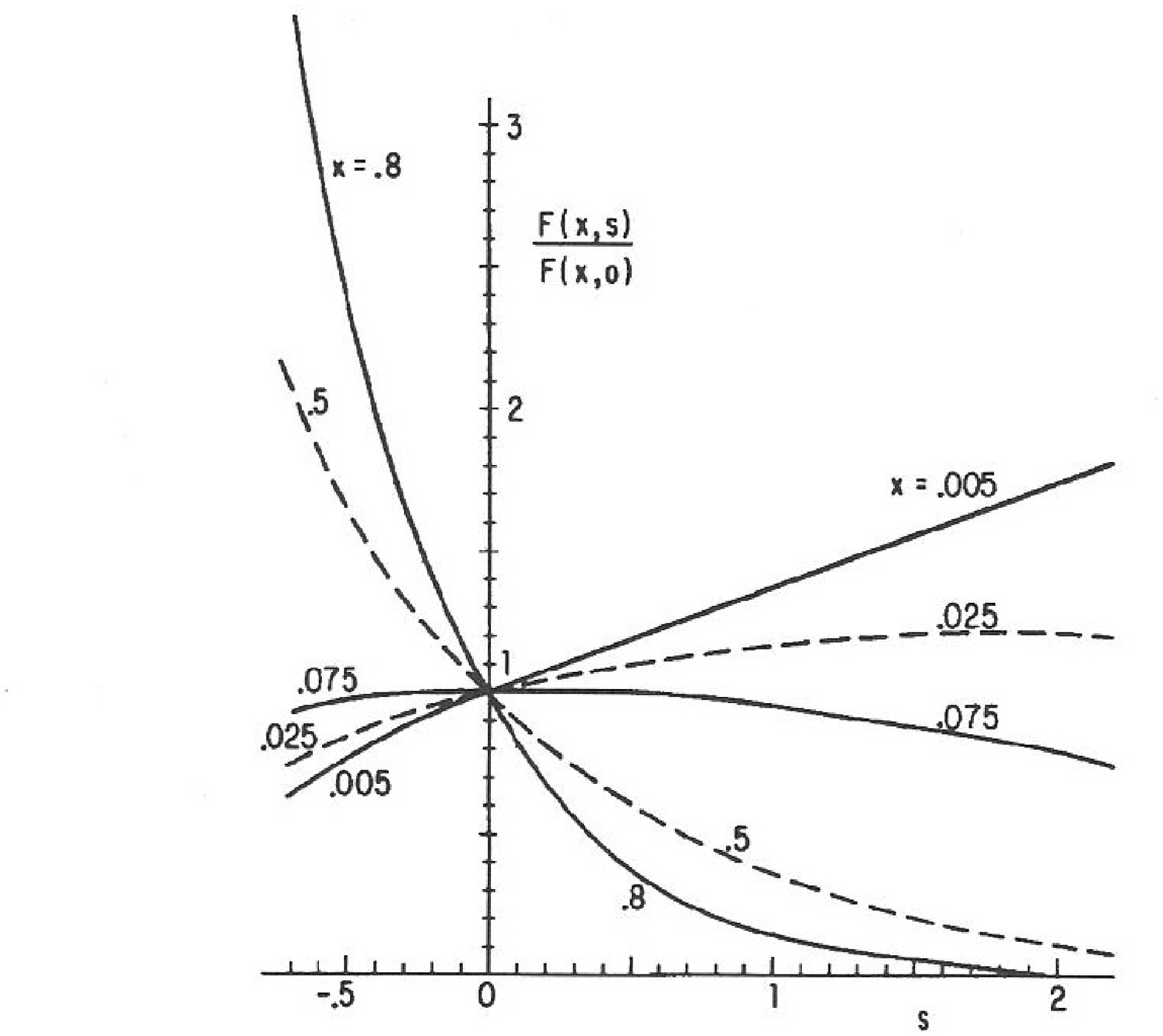,width=9cm}
\caption{Example of the evolution [in $Q^2$, or $s$ as in
Eq.~(1)] of a normalized
non-singlet structure function $F(x,s)/F(x,0)$ at fixed values of
$x$. The predicted trend has been corroborated in detail by a
multitude of experiments (and theorists).}
\label{Evolution}
\end{figure}

Perhaps
understandably, by our next paper  \cite{DGP1}, we were back to our peaceful
parted ways. Elaborating on work  \cite{Giorgio} by Giorgio Parisi (who had
Mellin-transformed the QCD results on $x$-moments into direct information
on the structure functions), we decoded their evolution at fixed $x$ for
varying $Q^2$. The result, shown in Fig.~(\ref{Evolution}), 
was to become heavily used... and
systematically referenced to authors of much later papers.
 While yowling, I plead guilty to having learned much later that
the simple underlying physics had been understood elsewhere: the
renormalization-group \cite{SP,GML}
 picture of seeing partons within partons was drawn by
Kogut and Susskind  \cite{KS}, the ``physical gauge'' diagrammatic image of
a parton dissociating into others is due to the usual Russian suspects
 \cite{Lip}, and the vintage QED analogue is nothing other than the
Weisz\"aker--Williams approximation  \cite{WW}.

During the times I have just described, it was my impression that
physics could not possibly be more hectic and interesting. I was wrong...

\subsection{The November Revolution}

Ten days of November 1974 shook the world of physics. 
Something wonderful and {\it almost}  \cite{AP} unexpected was to see the
light of day:  a  very discreetly charmed particle
 \cite{J,Psi}, a hadron so novel that it hardly looked like one.
Thirty years later, it is not easy to  recall  the 
collective ``high'' in which this discovery, and others to be made
in the two consecutive years, submerged the particle-physics community.
In my opinion, a detailed account that reflects well
the mood of the period is that by Riordan  \cite{Riordan}. In a
nutshell, the standard model arose from the ashes of the November
Revolution, while its competitors died honourably on the battleground.

In the early Fall of '74, Tom Appelquist and David Politzer 
had been looking leisurely at how asymptotic freedom could imply a
positronium-like structure for the $c\bar c$ bound states of a
charmed quark and its  anti-goodie-bean. In those days, both QCD and
charm were already fully ``established'' at Harvard, where one could,
without the scorn of one's peers, violate the first rule of scientific fairy tales:
not more than one fancy character per tale! Since Americans are often
short of vocabulary, my first contribution to the
subject was to baptize their toy {\it charmonium}. David and Tom's
first charmonium spectrum was so full of peaks and Coulomb-like that they
could not believe it themselves. They debated the problem long enough
for the experimental avalanche to catch up with them. It was  a heavy 
price to pay for probity.

Burt Richter was also caught in the avalanche. 
On a short visit to Harvard, and with a healthy disrespect for theory, Burt told
us that the electron spent some of its time as a hadron. 
In answer to a question by Tom, he explained that sufficiently narrow
resonances would escape detection in $e^+e^-$ colliders. Nobody around was aware of the possibility of catching the
devil by its radiative tail (the emission of photons by the colliding
particles widens the observed resonance on its $\sqrt{s}>M$ side). 
Our vain discussions came to an abrupt end; 
a rather urgent call summoning Burt back to SLAC delivered us from his 
scorn for theorists. Only in other
multi-world histories of our Quantum Universe  \cite{otroyo} do
charmed theorists get to talk also with Sam Ting, prior to the
Revolution\footnote{Sam did talk to a theorist at MIT, but Harvard's
charmed infatuation did not extend that far.}.

For an object of its mass, the $J/\psi$ is four orders of magnitude
narrower than a conventional hadron, or just one order of magnitude
wider than a weak intermediary (only because they are my good old
friends do some people escape reference and derision at this point).
 Of the multitude of theoretical papers that immediately hit the press
 \cite{ava}, only two attributed the narrow width to asymptotic
freedom, one by Tom and David  \cite{AP}, who had intuited the whole
thing before, the other by Sheldon Glashow and me  \cite{DeRG1}. I
recall Shelly storming the Lyman/Jefferson lab corridors with the notion of the
feeble three-gluon hadronic decay of the $J^P=1^-$ {\it
orthocharmonium} state, and I remember Tom and David muttering:
``Yeah''. Our paper still made it to the publishers in the auspicious
November, but only on the 27$^{\rm th}$, a whole week after the
article of our Harvard friends.

We did a lot in our extra week  \cite{DeRG1}. {\it Abusus non tollit
usum} (of asymptotic freedom) we related the hadronic width of the
$J/\psi$ to that of $\phi \to 3\,\pi$, to explain why the resonance
 {\bf had to be} so narrow. We correctly estimated the
yields of production of truly charmed particles in $e^+e^-$
annihilation, $\nu$-induced reactions, hadron collisions and
photoproduction. Our mass for the $D^*$ turned out to be $2.5 \% $
off, sorry about that. Re-reading this paper much later, I was surprised to see
that we had already discussed
mass splittings within multiplets of the
same quark constituency as {\it hyperfine}, a fertile notion. In
discussing paracharmonium ($J^P=0^-$) we asserted that {\it ``the
search for monochromatic $\gamma$'s should prove rewarding''}.
Finally, we predicted the existence of $\psi '$, but this time it was
our turn to be overtaken by the pace of discovery. Not every week of
my (scientific) life parallels this particular one. The commentary on
this paper in the book by Riordan \cite{Riordan} is one of those things
that my grandfathers would have liked to read. My grandmothers may
even have believed it!

\subsection{A First Year of Lean Cows}

At the time of the November Revolution, and for two or more years to follow,
there were hundreds of theorists ``out there'', determined to deride what is
now called the Standard Model. On rare occasions, this was due to healthy
scepticism or the invention of cute alternatives, e.g.~\cite{PZ}.
  More often than not the
reason was intellectual inertia or militant ignorance. But history is what
a society chooses to record, not what ``really'' happened. {\it It behooves
the wise to change their minds}, says a Spanish proverb, and now we all
sign onto the registry of wisdom.

I have relatively few vivid printable recollections of the times I am
discussing. One  concerns the late night in which the existence of
$P$-wave charmonia hit my head: we had been talking about $L=0$
states without realizing (we idiots!) that a bunch of $L=1$ charmonia
should lie between $J$ and $\psi '$ in mass. Too late to call Shelly
at home, I spent hours guessing  masses and estimating the obviously
all-important $\gamma$-ray transition rates. At a gentlemanly morning
hour I rushed on my bicycle to Shelly's office, literally all the way
in, and attempted to snow him with my findings. I was out of breath
and wits, the darned words would simply not come out. Shelly profited
to say: ``I know exactly what you are trying to tell me, there are
all these $P$-wave states etc., etc.'' He had figured it all out at
breakfast. I hated the guy's guts.

\begin{figure}[]
\hskip 3truecm
%\vspace*{2cm}
%\hspace*{-1.5cm}
\epsfig{file=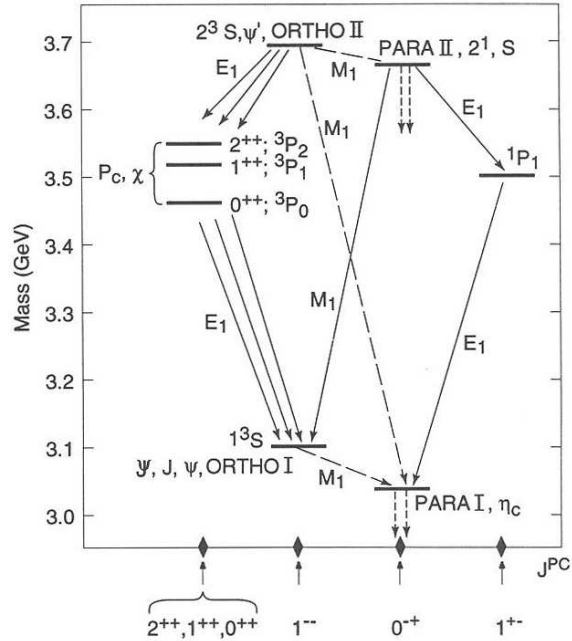,width=8.5cm}
\caption{ Harvard version of the masses and radiative transitions of
charmonium. All of these particles and most of their radiative decays were 
eventually observed.}
\label{Charmonium}
\end{figure}

In no time, David and Tom gathered forces with Shelly and me to produce
an article  \cite{ADGP} on {\it Charmonium spectroscopy.} Physical
Review Letters  was fighting its usual losing battle against
progress (in nomenclature, $\smile$) and did not accept the title. Our original
spectrum is reproduced in Fig.~(\ref{Charmonium}).

A few days after we submitted our opus  \cite{ADGP} 
we learned that a group at Cornell had been working on very
much the same subject (our usual competitors in Princeton
did not participate in the more phenomenological skirmishes
that were soon to take place). My recollection of how we learned about
Cornell's competition \cite{Erice} is not shared by Kurt Gottfried,
and I shall not reproduce it here. Our paper  \cite{ADGP} 
was a bit less elaborate (they'd say) or a bit more uncommitted to a particular
model (we'd say) than the Cornell paper but, I used to joke, the main difference 
between the two contiguous Physical Review
Letters  \cite{ADGP,Cornell} is that the Cornell analogue of 
Fig.~(\ref{Charmonium}) is mirror-imaged. In fact, their work was based
on explicit calculations with a confining linear-potential model, and
resulted in very successful predictions for the photon decay rates.

Early in 1975, Howard Georgi was back from his excursion into $SO(10)$
and $SU(5)$. He, Shelly, and I wrote a paper  \cite{DGG} whose style
reflects how high we rode. Here is how it began:

{\it Once upon a time, there was a controversy in particle physics.
There were some physicists$^1$ who denied the existence of structures
more elementary than hadrons, and searched for a self-consistent
interpretation wherein all hadron states, stable or resonant, were
equally elementary. Others$^2$, appalled by the teeming democracy of
hadrons, insisted on the existence of a small number of fundamental
constituents and a simple underlying force law. In terms of these
more fundamental things, hadron spectroscopy should be qualitatively
described and essentially understood just as are atomic and nuclear
physics.}

We proceeded to do just that. Our second reference
was to Gell-Mann and Zweig for the quark model
\cite{GMQuark,Zweig}.
Like most people to date, we did not know at the
time that Andr\'e Petermann may have been the first to consider
spinor constituents of hadrons with ``constituent masses" and
fractional charges \cite{Andre}. I cannot resist quoting him:
{\it Ou bien, si l'on veut pr\'eserver la conservation de la charge,
ce qui est hautement souhaitable, les particules doivent alors
avoir des valeurs non-enti\`eres de la charge. Ce fait est d\'eplaisant
mais ne peut, apr\`es tout, \^etre exclu sur des bases physiques.}
To the non-relativistic version of the quark model, Howard, Shelly and I
added chromodynamic quark interactions
entirely analogous to their electrodynamic counterparts. To this day
it is unclear why the ensuing predictions are so good. Our paradigmatic
result was the explanation of the origin and magnitude of the
$\Sigma$--$\Lambda$ mass difference. The two particles have the same
spin and quark constituency, their mass difference is a {\it
hyperfine} splitting induced by spin--spin interactions between
the constituent quarks.
A little later, the ``MIT bag" community published their 
bag-model relativistic
version  \cite{MIT} of the same work.

In  \cite{DGG} we also predicted the masses of all ground-state
charmed mesons and baryons and (me too, I'm getting bored with this)
we got them right on the mark. Competing predictions based on an
incredible $SU(4)$ version of the Gell-Mann--Okubo $SU(3)$ mass formula,
and also the more sensible bag results, turned out to be wrong.
Interestingly, only one person trusted a ``QCD-improved'' constituent
quark model early enough to make predictions somewhat
akin to ours: Andrei Sakharov  \cite{Sach}.

\subsection{Good News at Last}

While theorists faithfully ground out the phenomenology of QCD,
experimentalists persistently failed to find decisive signatures of our
Trojan horse: the charmed quark. At one point,  the upper limits on the
$\gamma$-ray transitions of charmonia were well below the theoretical
expectations. Half of the $e^+e^-$ cross-section above $\sqrt{s}=4$ GeV was
due to charm production, said we. Who would believe that experimentalists
couldn't tell?

In the winter of '75 we saw a lone ray of light. As Nick Samios
recalls in detail in \cite{NSamios}, 
a Brookhaven bubble-chamber group  \cite{Nick} pictured
a $\Delta S=-\Delta Q$ event, forbidden in a charmless world, and
compatible with the chain:
\begin{eqnarray}
&&\nu_\mu\;p  \rightarrow \Sigma^{++}_c \mu^-\nonumber\\
&&
\Sigma^{++}_c  \rightarrow \Lambda^+_c\; \pi^+\nonumber\\
&&
\Lambda^+_c  \rightarrow \Lambda\;\pi^+\;\pi^+\;\pi^- \; .\nonumber
\end{eqnarray}
The mass and mass difference of the two charmed baryons caught here
in one shot sat on top of the predictions  \cite{DGG}. This was a
source of delight not only for us, but also for the experimentalists
involved. 

In the summer of '75 ---after a year of upper limits incompatible with
the theoretical expectations---
the first evidence finally arose for the $P$-wave
states of charmonium  \cite{Pc}. The DESY group did
not refer to  the theorists who suggested their search; they are hereby
punished: they do not get a  reference, and they will remain
eternally ignorant of my juicy version of this story, concerning their
competition with SLAC.

The discovery of a positronium-like $c\bar c$ spectrum converted
crowds of infidels to the quarker faith. And the charmed quark was
to continue playing a crucial role in the development and general
acceptance of the standard lore.

\subsection{Yet Another Year of Lank Cows}

Measurements at SLAC of the $e^+e^-$ cross section into hadrons  \cite{Sig}
showed a doubling of the yield and structure aplenty as the $\sqrt{s}=4/5$
GeV region is crossed. Much of the jump {\bf had} to be due to the
production of charmed pairs, which were not found. Howard
Georgi and I innocently
believed that a sharpening of the arguments would help.

In the space-like domain, $s\!<\!0$, QCD predictions 
for $e^+e^-$ annihilation are insensitive
to thresholds, bound-state singularities and hadronization caveats.
For years, theorists had been unjustifiably applying the predictions
to the time-like domain wherein experimentalists insist on taking
$e^+e^-$ data. In a paper  \cite{HYo} whose rhythmic title {\it
Finding Fancy Flavours Counting Colored Quarks} was duly censored, we
transferred the $e^+e^-$ data, via a dispersion relation, to a
theoretically safer space-like haven.  This somersault  \cite{Adler}
allowed us to conclude that {\it the old theory with no
charm is excluded, the standard model with charm is acceptable if
heavy leptons are produced, and six quark models are viable if no
heavy leptons are produced}. Thus, anybody listening to the other
voice in the desert (that of Martin Perl, who was busy demonstrating
that he had discovered the $\tau$) had no choice but 
charm. 

Our work was improved by Poggio, Quinn and Weinberg  \cite{PQW}, who
realized that one could, in the complex $s$-plane, work in a contour
around the real axis where perturbative QCD can still be trusted,
whilst the distance from the dirty details of real life is judged
safe. The imprimatur of Steve {\it et al.,} attracted considerable
attention to this aspect of QCD and strengthened our conclusion: the
measured total cross section, analysed on firm theoretical grounds,
implied the existence of charm and of a new heavy lepton.

To come back to the role of charm in QCD, the determination of
$\Lambda$ had been one of my fixations  \cite{yo} since the birth of
this unique QCD parameter. In the paper with Howard  \cite{HYo} we
made an attempt to get a meaningful result, using fresh
(but incomplete) data on deviations from scaling in deep inelastic
scattering of muons on iron  \cite{WC}. We concluded {\it The
$Q^2$-dependence for $\Lambda=.32$ GeV agrees qualitatively with the
$\mu$-Fe data... If the standard model with one heavy lepton is
correct, the actual value of $\Lambda$ is probably between 1 and .32
GeV}. The announced lepton was the $\tau$.

\subsection{Charm is Found}

No amount of published information can
compete with a few minutes of conversation. The story, whose moral
that was, is well known. For the record, I should tell it once more
 \cite{Aachen,Goldhaber}:

Shelly Glashow happened to chat with  Gerson Goldhaber in an airplane.
 Something unusual took place. The East Coast theorist
managed to convince the West Coast experimentalist of something.
There was no way to understand the data unless charmed particles were
being copiously produced above $\sqrt{s}=3.7$ GeV. The
experimentalists devised an improved (probabilistic) way to tell
kaons from pions. In a record 18 days two complementary SLAC/LBL
subgroups found convincing evidence for a new particle  
with all the earmarks of charm \cite{Gerson}. The charmonium
advocates at Cornell had been trying for a long time to convince the
experimentalists to attempt to discover charm by sitting on the
$\Psi(3440)$ resonance, or on what would become a ``charm factory":
$\Psi(3770)$ \cite{Corncharm}. Alas, they failed.

The observation of charmed mesons
ought to have been the happy ending, but there was a last-minute
delay. The invariant-mass spectrum of recoiling stuff in $e^+e^-\to
D^0\,(D^\pm)+...$ had a lot of intriguing structure, but no clear
peak corresponding to $D^0\bar D^0$  \cite{Gerson} or $D^+D^-$
 \cite{Peru} associated production. Enemies of the people rushed to
the conclusion that what was being found was an awful mess, and not
something as simple as charm.

\begin{figure}[]
\hskip 1.0truecm
%\vspace*{2cm}
%\hspace*{-1.5cm}
\epsfig{file=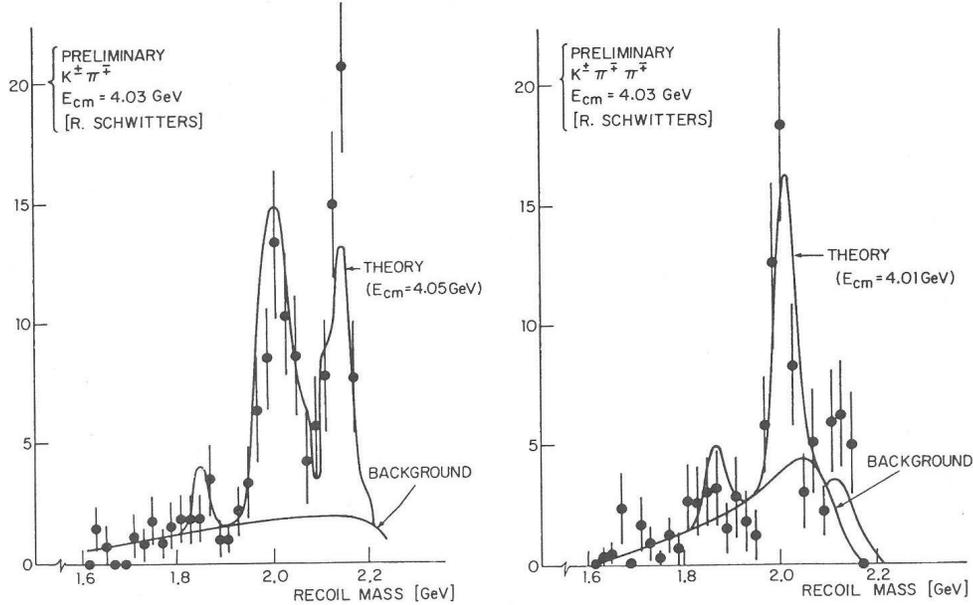,width=13.5cm}
\caption{Predicted and observed invariant-mass spectra, recoiling
against neutral and charged $D$'s. The theoretical
curves are a one-parameter fit.}
\label{Charm}
\end{figure}

But we had one  last unspent cartridge  \cite{Charm}. We expected $D\bar
D$, $D\bar D^*+\bar D D^*$, and $D^*\bar D^*$ production to occur in
the ``spin'' ratio 1:4:7 (thus the $D\bar D$ suppression). We trusted
our prediction  \cite{DGP1} $m(D^*)-m(D)\simeq m(\pi)$, which implies
that for charm production close to threshold, the decay pions are
slow and may be associated with the ``wrong'' $D$ or $D^*$ to produce
fake peaks in recoiling mass. Finally, we knew that the charged $D$'s
and $D^*$'s ought to be a little heavier than their neutral sisters.
The $D^*$ decays had to be very peculiar: $D^{*0}\to D^+\pi^-$ is
forbidden, $D^{*0}\to D^0\gamma$ competes with $D^{*0}\to D^0\pi^0$,
etc. On the basis of these considerations (and with only one fit
parameter) we constructed the recoil spectra shown in 
Fig.~(\ref{Charm}). Case closed!

``Enemies'', ``cartridge''... Why am I using such  belligerent terms?
Not only because of the vehement competition between divers
schools of theorists. I have a keen memory of the East Coast
experimentalists with whom I talked at the time (not a significant
sample, in the statistical sense). They were strongly motivated to disprove all
theories and theorists, perhaps permeated by some arcane Californian
faith that nature is intrinsically unfathomable. They most
certainly did not have an instruction on their data-analysis program
stating IF (RESULT = STANDARD) THEN STOP. This made life
most enjoyable and the case for the standard model {\it veeeery} strong.
Having worked over the last few years in astrophysics, I have discovered
to my amazement that challenging the corresponding standard models is
{\bf not} a main motivation of theorists and 
observers, quite the contrary, see e.g.~\cite{GRB}.

\subsection{Back to the Deep Inelastic Domain}

\begin{figure}[]
\vspace*{-2cm}
%\hskip 0.5truecm
%\hspace*{-1.5cm}
\epsfig{file=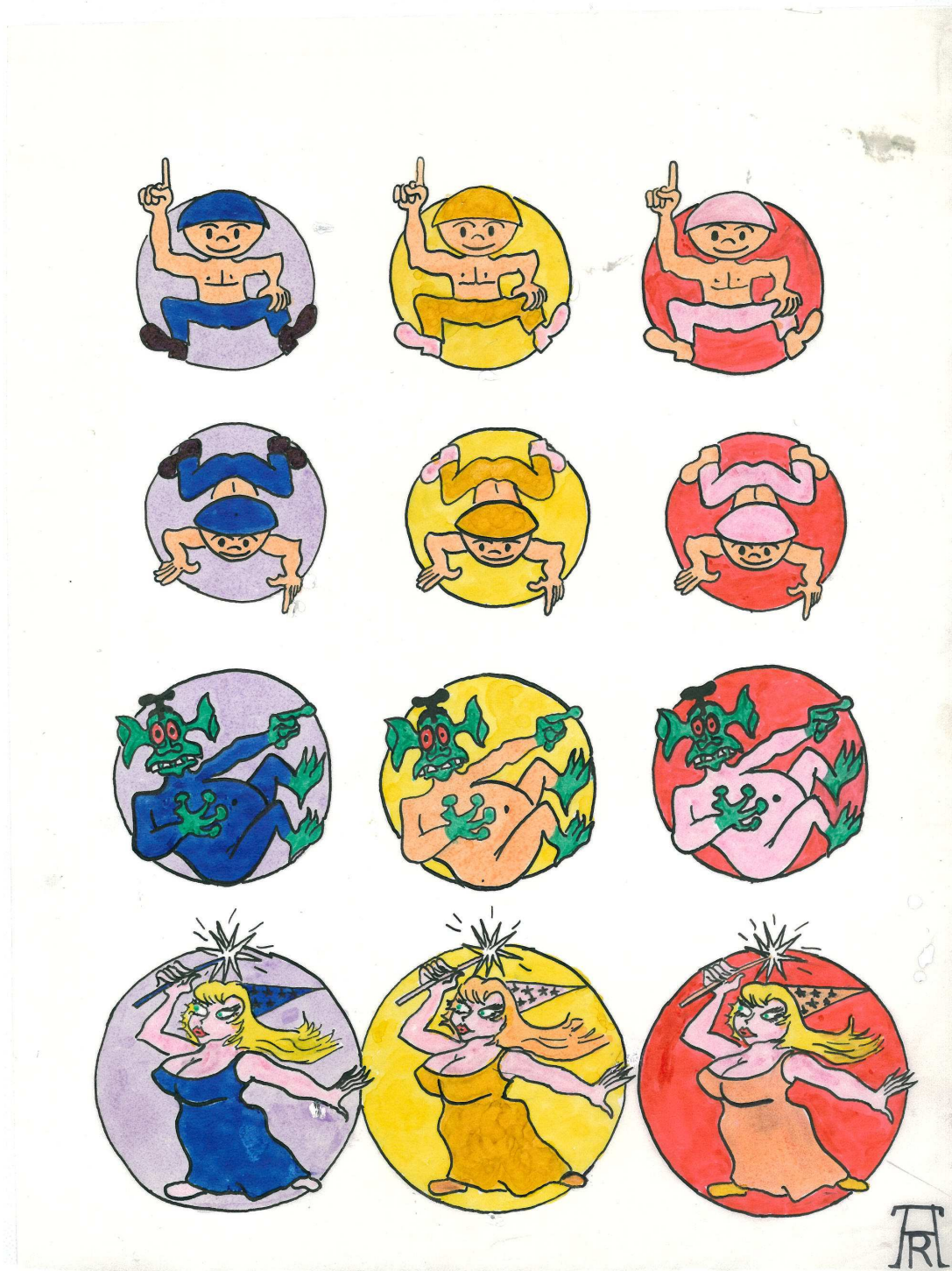,width=14.7cm}
\caption{The first 4 quark flavours in their 3 colours (those of the
Spanish republican flag), {\it c.} 1974.  Understandably, I have refrained from
updating this figure to include {\it top} and {\it bottom}. I would
have done it, if the names I proposed at the time for their flavours
({\it tenderness} and {\it beauty}) had caught.}
\label{Quarks}
\end{figure}

By the summer of '76 the skeleton of the standard model was almost
complete: charm was established and  the $\tau$-lepton  
\cite{MartinPerl} was
making its unhasty way towards acceptance.  A talk I gave that summer
\cite{TbilisiYo} 
---of which Fig.~(\ref{Quarks}) was a transparency---
could have been entitled {\it Status of the Standard Model},
even if it was not yet {\it standard}, or 
called that way. The masses of some of
the missing bones ($b$ and $t$, and the various neutrinos),
but not those of others ($W$ and $Z$) were still utterly
unknown, and the Yang--Mills structure of the theory was essentially untested.
Yet, a soon as the ``sequential" pair $(\tau,\,\nu_\tau$) was established,
``we all knew'' that $b$ and $t$ had to exist, not only to cancel anomalies
\cite{BIM}, but to fit into a decent theory of CP violation \cite{KoMa}.

One reason why QCD was not better tested is that the only precise
deep inelastic data pertained to modest momentum transfers. An
analysis in terms of scaling, and QCD deviations thereof, would have
to face tedious technicalities (target-mass corrections and
higher-twist effects) and flagrant departures from scaling (prominent
nucleon-excitation resonances). Though all of this squalor was well
beneath the proverbial dignity of a self-respecting theorist, David,
Howard and I decided to face the difficulties, in a paper
unpronounceably entitled {\it Demythification of Electroproduction
Local Duality and Precocious Scaling} and in its accompanying letter
 \cite{DGP2}.

I believe {\it ``Demyth''} to be one of the nicest productions in
which I have participated. I suspect it is the most influential, but
I {\bf know} (thanks to currently available electronic means) that it
is not very much quoted.
One reason why I think {\it Demyth} was influential is that one of
its ingredients, the performance of target-mass corrections via
$\xi$-scaling  \cite{Otto,DaHo} provoked a violent reaction from
numerous groups  \cite{antixi}. We believed theirs to be a
misunderstanding, but to this day I am not sure whether it was our
response  \cite{DGP4}, or something else, that defused this
altercation  \cite{play}.

\begin{figure}[]
\hskip 0.1truecm
%\vspace*{2cm}
%\hspace*{-1.5cm}
\epsfig{file=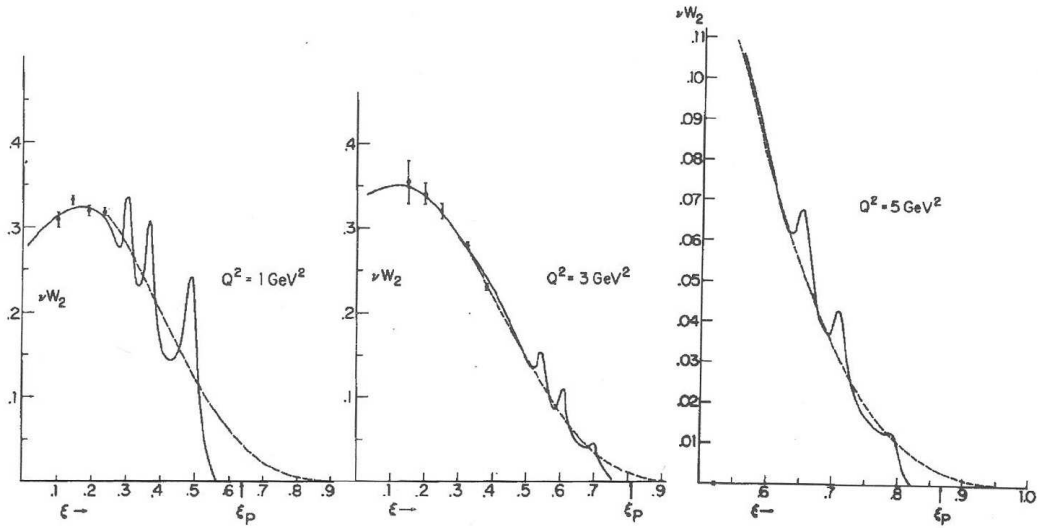,width=14.5cm}
\caption{Bloom--Gilman duality. The smooth curves are QCD-evolved
(to leading twist) from data at higher $Q^2$. The curves with resonances
are fits to the large-$\xi$ data, at various fixed values of $Q^2$.}
\label{Duality}
\end{figure}

Mysterious dualities are a recurring theme in physics. Bloom--Gilman
duality is the observation that at low $Q^2$ a structure function
shows prominent nucleon resonances, which ``average'' to the
``scaling'' function measured at some higher $Q_0^2$, and slide down
its slope as $Q^2$ varies, all as in Fig.~(\ref{Duality}).

We argued in  \cite{DGP2} that this duality is a consequence of QCD,
inevitable if scaling is ``precocious,'' as it must be for small
$\Lambda$ (a fraction of a GeV). In Fig.~(\ref{Duality}) the proton-pole positions
are at the places labelled $\xi_p$. The area under their $Q^2$-dependent
contributions ($\delta$ functions in $\xi$) approximately equal the
areas under the corresponding smooth curves at large $\xi$. 
This justifies a posteriori
the analysis of \cite{yo,GT} and explains why those early attempts 
at QCD phenomenology resulted in a reasonable value for $\Lambda$.
It is a satisfaction, particularly for theorists, to be the first to ``measure"
a fundamental constant of nature.

With the benefit of hindsight, I view as the most interesting result
of  \cite{DGP2} the conclusion that, in the comparison of experiment
and QCD theory, {\it excellent agreement is obtained using}
$\alpha_s(Q^2=4\;{\rm GeV}^2)/\pi=0.17$. This corresponds to
$\Lambda\sim 0.5$ GeV and is, I believe, the first {\it solid}
determination of this basic constant.

The consensus that the observed scaling deviations smelled of QCD was
not triggered by theorists, but by an analysis of neutrino data by
Don Perkins {\it et al.}  \cite{Don}. This test of QCD was not very
severe. One reason is the neglect of higher twists  \cite{LM,play}.
Furthermore, it is not easy to measure the incident
neutrino energy. Thus, in an unintended implementation of Bloom--Gilman
duality, $F(x,Q^2)$ is significantly blurred in $x$ and $Q^2$. At the
low $Q^2$ of a good fraction of the data, there 
must be bumps such as those of
Fig.~(\ref{Duality}). Had the resolution of neutrino experiments
been as good as that of
their electron-scattering counterparts, the bumps would have been
visible, and the data analysis would have had to be quite different.

With a pinch of poetic licence one could assert that, early on, many
concluded that QCD was quite precise, but only  because the data were not.

\subsection{Seeing is Believing}

Quarks have not been seen and may  \cite{DGJ} never be. But their
manifestation as quark jets was apparent since 1976  \cite{Qjets}.
Kogut and Susskind  \cite{KS} argued that the gluon, an important
character that I have not yet discussed explicitly, could show up in
the same way: the elementary process $e^+e^-\to \bar q\, q\, g$ may
result in three-jet final states.

Further work on QCD jets was often based on ``intuitive perturbation
theory'', an appellation perhaps meant to reflect a fundamental
lack of understanding. Decorum was regained by
the work of Sterman and Weinberg  \cite{SW}, Georgi and Machacek
 \cite{HoMa} and Farhi  \cite{EF}, who exploited the fact that in QCD,
as in QED, there are ``infrared-safe'' predictions  \cite{HDP}, not
sensitive to the long-distance dynamics that, in QCD, are intractable
in perturbation theory.

\begin{figure}[]
\hskip 1.5truecm
%\vspace*{2cm}
%\hspace*{-1.5cm}
\epsfig{file=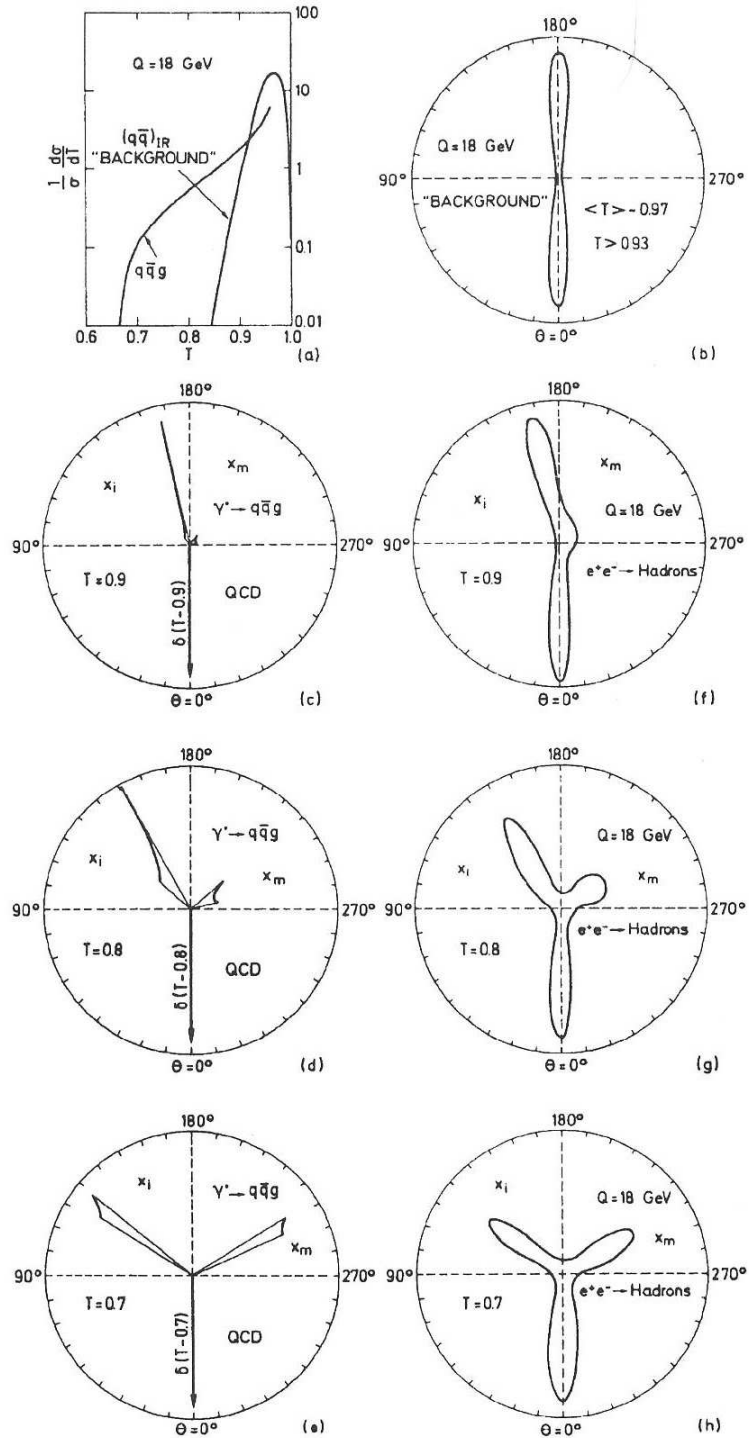,width=11.5cm}
\caption{Examples of predictions  for three-jet
distributions at different fixed thrusts, $T$. The left column contains
the leading perturbative QCD predictions; in the right one the
results are smoothed for ``hadronization" effects. More often than not,
the small jet is produced by a gluon.}
\label{Gluon}
\end{figure}

One infrared-safe observable is the ``antenna'' pattern of energy
flow in an ensemble of hadronic final states in $e^+e^-$
annihilation, properly reoriented event by event to compensate for
the vicious quantum mechanical penchant for uncertainty.  We foretold
 \cite{DEFG} the pattern, binned in ``thrust''  \cite{EF},  to be that
of Fig.~(\ref{Gluon}). This three-jet structure and the QCD-predicted details of
the angular or energy distributions played an important role in the
``discovery of the gluon'', a subject whose denouement 
(that gluons are for real) has been
described in detail by James Branson in \cite{JB}. 

\section{Grand Unification and proton decay}

I remember an occasion, in 1974, when I was visiting SLAC.
``Bj" Bj{\o}rken told me that he had recently read the article on $SU(5)$ 
 by Georgi and Glashow \cite{SU5}. 
He said that, as he was reading it, his heart was pounding faster and faster
at the beauty of the unification of QCD and the electroweak theory,
the convincing explanation of the commensurate integer and fractional
charges of quarks and leptons, etc., etc. But, Bj said, his heart sunk when he 
read that protons were not forever. At the time I thought his reaction was
precisely the wrong one. Proton decay, I thought, was the best part
of this Grand Unification: a candidate for the ultimate YM theory, including
such a fantastic prediction! With the benefit of thirty years of hindsight,
I am beginning to wonder whether Bj was right. I may have done this
before, for Howard, Shelly and I once wrote a note \cite{Trini} on an 
alternative {\it trinification} model, based on the gauge group
$SU(3)\otimes SU(3)\otimes SU(3)\otimes Z_3$, with all the virtues
of $SU(5)$ (or $SO(10)$ \cite{SO10}), but no gauge-mediated proton decay.

\begin{figure}[]
\hskip 1.5truecm
%\vspace*{-1cm}
%\hspace*{-1.5cm}
\epsfig{file=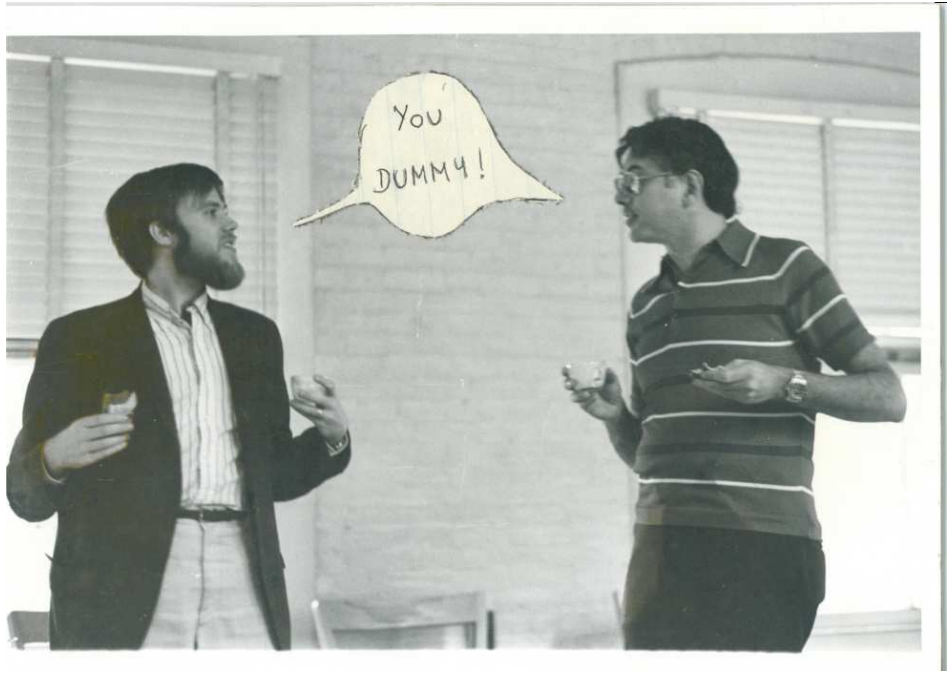,width=12.5cm,angle=0}
\caption{Howard Georgi and Shelly Glashow discuss Grand Unified
Models and eat their cake too.}
\label{HS}
\end{figure}

Georgi, Hellen Quinn and Weinberg  \cite{GQW} were first to compute the 
renormalization
of the gauge couplings of the $SU(3)$ and $SU(2)\otimes U(1)$ subgroups
of a Grand Unified model, thereby refining the postdiction for the weak
mixing angle and predicting the lifetime of the proton. In a sense,
they also predicted, this time correctly ---but not overtly--- the mass of
the proton: some three times $\Lambda$, the ``infrared"
momentum scale at
which the QCD coupling becomes intractably large\footnote{I have
learned this way of looking at things from Savas Dimopoulos.}.
John Ellis, Mary K. Gaillard and collaborators were the first to dare
extend these renormalized strictures to relations between fermion
masses \cite{JohnMary}. The graph showing the unifying coupling
constants is so famous that I shall not show it here, not even in its
(still tenable) MSSM extension (the first M, I have always contended,
describes the credibility of the {\bf Minimal} Supersymmetric extension
of the Standard Model). Instead of this graph, I show in Fig.~(\ref{HS}) a
1974 picture of Howard and Shelly debating the advantages or
disadvantages of $SO(10)$ {\it versus} $SU(5)$.

Grand unifications of YM theories and proton instability
 actually slightly preceded the work in
 \cite{SU5}. Indeed, Jogesh Pati and Abdus Salam were the first
 to introduce the elegant idea of lepton number as a fourth colour
 \cite{Jogesh}. I learned from Bram Pais, via Georgi \cite{HGeorgi},
 that their $SU(2)\otimes SU(2)\otimes SU(4)$ group even contained the full 
 $SU(3)\otimes SU(2)\otimes U(1)$  gauge structure of the standard
 model! But they proceeded to break colour $SU(3)$ into integrally-charged
 Han--Nambu quarks, thereby distantiating themselves 
 from the current standard lore.
 
 The Harvard theorists were, this time, capable of immediately
 convincing experimentalists
 that proton decay was interesting to look for. Larry Sulak, in particular, was
 enthusiastic with the idea, and he was among those who
 developed the inverse-osmosis technique
 that proved to be crucial in making water 
 ---at a reasonable cost---
 sufficiently transparent to exploit
 a large water-\v Cerenkov detector. Larry deprived his colleagues at the University
 of Michigan of an elevator, to install in its shaft
 a sufficiently tall water container and test the technique. His colleagues
 may not have appreciated their extra stair-climbing efforts,
 but eventually this kind of detectors made great serendipitous discoveries:
 neutrinos from SN1987A, neutrino oscillations,~...
 Proton decay has not been observed to date, and who knows what future
proton-decay detectors may uncover.

\section{Gravity}

Unbeknown to most, the Apollo 11 astronauts actually did something
useful, other than testing Moon boots. In 1969, they placed the first 
passive laser reflector on the Moon. After so many years of sporadic
lunar ranging measurements, some of the length parameters
describing the lunar orbit are known with millimetre precision.
In a considerable improvement of Galileo's supposed experiment
at the Leaning Tower of Pisa, the 
Earth and the Moon are measured to ``fall'' towards the
Sun with the same acceleration, with a precision of 
$\sim 2\times 10^{-13}$ \cite{KN1}.

The gravitational self-mass of a uniform extensive body is 
$\Delta M\propto G_N\, M^2/R$. More precisely, this quantity
is $\Delta M_\otimes\approx -\, 4.6\times 10^{-10}\;M_\otimes$
for our planet, and $\Delta M_\mu\approx -\, 0.2\times 10^{-10}\;M_\mu$
for its satellite. If these self-mass contributions were attracted by the Sun
differently from the bulk of the mass of these two bodies, differing
accelerations would result \cite{KN2}. In actual numbers, we know
that the Earth's bulk acceleration and that of its self-gravity are
equal to a precision of $2\times 10^{-3}$. 

\begin{figure}[]
\hskip 3truecm
%\vspace*{2cm}
%\hspace*{-1.5cm}
\epsfig{file=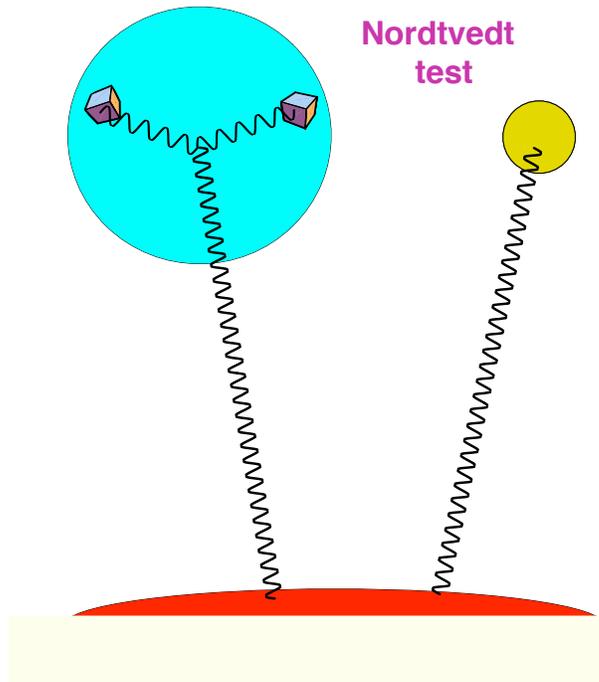,width=8cm}
\caption{The Earth and the Moon being attracted to the Sun. The triple-graviton
vertex represents the pull by the Sun on the gravitational self-energy
of the Earth.}
\label{Nor2}
\end{figure}

As I recalled in the Introduction, Einstein's gravity is akin to a
 non-Abelian YM theory in the sense that gravitons gravitate. 
 The diagrammatic translation of the previous paragraph is
 shown in Fig.~(\ref{Nor2}). What all this means is that we know the
 triple-graviton coupling to be what it should be, to 2 thousands.
This is better than the precision to which we know the 
``triple-gauge" couplings of intermediate vector bosons, or
the equality (up to group-theory factors) of the colour charges
of quarks and gluons, which is hard to measure ``directly"
\cite{Beni}. There is a long way to go before we have tested YM 
vector-boson theories to an astronomically satisfactory  precision!

\section{Afterword}

The most profound mystery of nature is that it can be
pictured and tamed in mathematical terms; the deepest quality of
scientists is that they can imagine conceptual abstractions that turn
out to be measurable physical realities. Yang--Mills theories are
an example, and the visionary confidence of some of their
developers always makes me recall the very same old story:

My fabulous math teacher in high school was
totally embedded in math, in the sense that he fully believed
in the actual reality of mathematical abstractions. He would go to
the blackboard and draw an $X$ and a $Y$ axis. From the origin of
coordinates his pinched fingers would slowly deploy a fictional $Z$
axis, pointing to the middle of his mesmerized audience. And for the
rest of his lecture, as he paced up and down, he would never forget
to hold an imaginary line with a steady hand, as he bent to cross
under the $Z$ axis.
\vskip .2cm
{\bf Apologies \& Acknowledgements.} Surely I must have missed 
some important references, and I am sorry for that.
I thank Gerard 't Hooft for giving me the
opportunity to recall these good old times, and all my collaborators
---named and unnamed--- for having put up with me for years.

\end{document}